\documentclass[twocolumn,amsmath,amssymb,showpacs,superscriptaddress,floatfix]{revtex4}

\usepackage{graphicx}% Include figure files
\usepackage{dcolumn}% Align table columns on decimal point
\usepackage{bm}% bold math

\begin{document}

\title{Radio-frequency dressed state potentials for neutral atoms}

\author{S.~Hofferberth}
 \email{hofferberth@atomchip.org}
 \affiliation{Physikalisches Institut, Universit\"at Heidelberg, D-69120 Heidelberg, Germany}
\author{I.~Lesanovsky}
 \affiliation{Physikalisches Institut, Universit\"at Heidelberg, D-69120 Heidelberg, Germany}
 \affiliation{Institute of Electronic Structure and Laser, Foundation for Research and Technology,GR-71110 Heraklion, Greece}
\author{B.~Fischer}
\author{J.~Verdu}
 \affiliation{Physikalisches Institut, Universit\"at Heidelberg, D-69120 Heidelberg, Germany}
\author{J.~Schmiedmayer}
 \affiliation{Physikalisches Institut, Universit\"at Heidelberg, D-69120 Heidelberg, Germany}
 \affiliation{Atominstitut \"Osterreichischer Universit\"aten, TU-Wien, Vienna, Austria}

\date{\today}
\pacs{39.90.+d, 03.75.Be}

\begin{abstract} Potentials for atoms can be created by external fields
acting on properties like magnetic moment, charge, polarizability,
or by oscillating fields which couple internal states. The most
prominent realization of the latter is the optical dipole
potential formed by coupling ground and electronically excited
states of an atom with light. Here we present an
experimental investigation of the remarkable properties of
potentials derived from radio-frequency (RF) coupling between
electronic ground states. The coupling is magnetic and the vector
character allows to design state dependent potential landscapes.
On atom chips this enables robust coherent atom manipulation on
much smaller spatial scales than possible with static fields
alone. We find no additional heating or collisional loss up to
densities approaching $10^{15}$ atoms / cm$^3$ compared to static magnetic traps. We demonstrate the creation of Bose-Einstein condensates in RF potentials and investigate the difference in the interference between two independently created and two coherently split condensates in identical traps. All together this makes RF dressing a powerful new tool for micro manipulation of atomic and molecular systems.
\end{abstract}

\maketitle

Dressing of internal states of an atom with an external field is a well known technique in quantum optics \cite{Cohen92}. The coupling of atomic states to an oscillating field leads to new eigenstates and eigenenergies in the combined system. These dressed states can form adiabatic potentials, which can be employed for atom trapping and manipulation. The most prominent example is the optical dipole potential \cite{Grimm00}
created when intense coherent light couples ground and
electronically excited states of an atom. To create conservative
potentials for coherent manipulation spontaneous relaxation of the
excited state has to be avoided and hence the light field has to
be far detuned. Consequently the magnitude and shape of the dipole
potential is given by the local intensity of the light field. Such
far detuned dipole potentials are widely used in ultra cold atom experiments \cite{atomtraps}.

Dressing can also be achieved by coupling hyperfine components of
the electronic ground state by a magnetic radio-frequency (RF) or
micro-wave (MW) field. Dressed state potentials resulting from RF
coupling of two spin states in a magnetic field have been studied
in neutron optics \cite{Muska87}. Adiabatic potentials induced by
coupling hyperfine states with a micro wave have been proposed in Ref. \cite{Agost89}, and a detuned micro-wave has been used
for trapping ultra cold Cs atoms \cite{Spree94}. The trapping of
neutral atoms with RF induced potentials was proposed in
\cite{Zobay01} and first demonstrated for thermal Rb atoms
\cite{Colom04}. Recently RF dressed state potentials were employed
for coherent splitting of a one-dimensional Bose-Einstein
condensate and matter-wave interference \cite{Schum05}. In this paper, we give for the first time a full experimental demonstration and analysis of the remarkable properties of these adiabatic RF
dressed state potentials, which make them a versatile new tool for
experiments with ultra cold atomic and molecular systems.

\section{RF dressed state potentials}

In contrast to the optical dipole potential only electronic ground
states are involved in magnetic RF dressing. Consequently there are
no spontaneous relaxation processes and conservative potentials can
also be created with on-resonant radiation. The coupled states are
the magnetic sub-states of an atomic hyperfine ground state and a
spatially dependent Zeeman-shift can be used to locally change the
coupling strength. Furthermore, the coupling is magnetic and of
vector nature, hence not only the amplitudes but also the (local)
orientation of the involved RF and static magnetic fields
determine the coupling strength \cite{Lesan06}.

For our experiments we create the RF dressed state potentials by
combining a static Ioffe magnetic trap
($\mathbf{B}_S(\mathbf{r})$) with a two component homogeneous
RF-field ($\mathbf{B}_\textrm{RF}$) with frequency
$\omega_\textrm{RF}$ and relative phase shift $\delta$.
\begin{eqnarray}
    \mathbf{B}_S(\mathbf{r}) &=& Gx\mathbf{e}_x-Gy\mathbf{e}_y+B_I\mathbf{e}_z   \\
    \mathbf{B}_\textrm{RF} &=& B_{\textrm{A}}\mathbf{e}_x   \cos (\omega_\textrm{RF} t)\nonumber
    \\ &&+ B_{\textrm{B}} \mathbf{e}_y   \cos (\omega_\textrm{RF} t + \delta
    ).
  \label{Eq:Bfields}
\end{eqnarray}
$G$ is the gradient of the underlying static quadrupol field and
$B_\text{I}$  the magnitude of the Ioffe field. Following
\cite{Lesan06} the adiabatic RF-potential created can be written as
\begin{eqnarray}
  V = m_F g_F \mu_B \sqrt{ \Delta(\mathbf{r}) ^2 +
  \Omega_\textrm{RF}(\mathbf{r}) ^2}.
  \label{Eq:ring_dw}
\end{eqnarray}
with
\begin{eqnarray}
    \label{Eq:ring_dw2}
    \Delta(\mathbf{r}) &=& \left|\mathbf{B}_S(\mathbf{r})\right|-\frac{\hbar\omega}{|g_F  \mu_B|}\\
    \frac{8\left|\mathbf{B}_S(\mathbf{r})\right|^2\Omega^2_\textrm{RF}(\mathbf{r})}{B_{\textrm{A}} ^2 + B_{\textrm{B}} ^2} &=&
        2B_\text{I}\left[B_\text{I}+\left|\mathbf{B}_S(\mathbf{r})\right|\sin(2\alpha)\sin\gamma\right]  \nonumber \\
        & & + G^2\rho^2\left[1-\cos(2\alpha)\cos(2\phi)\right.
        \\& &
        \left.+\sin(2\alpha)\sin(2\phi)\cos\gamma\right].\nonumber
\end{eqnarray}
where $\rho=\sqrt{x^2+y^2}$, $\phi=\arctan{\frac{y}{x}}$ are polar
coordinates, $\tan(\alpha) =\frac{B_B}{B_A}$, and $\gamma = -
\frac{g_F}{\left|g_F\right|}\delta$ is the effective phase shift
which depends on the $g$-factor $g_F$ of the hyperfine manifold
under consideration.

\section{Implementation on an Atom Chip}

\begin{figure}
    \includegraphics[angle=0,width=8cm]{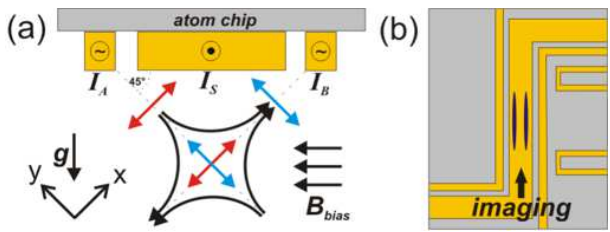}
    \caption{Atom chip setup for RF-potentials.
    (a) Side view of the three-wire setup. A broad ($100$ $\mu$m)
    central wire creates the static trapping potential.
    Smaller ($10$ $\mu$m) wires on each side provide the
    oscillating magnetic fields that create the adiabatic RF-potentials. The static trap is positioned so that the two RF-fields are perpendicular at its center.
    (b) Top view of the atom chip showing the relevant wires.
    The longitudinal confinement of the static trap is provided
    by the outer leads of the Z-wire. Additional U-wires
    on the side can be used to increase this confinement.}
    \label{Fig:Setup}
\end{figure}

For the experimental realization of RF dressed state potentials,
atoms confined in a static magnetic trap have to be irradiated
with a radio-frequency field. Atom chips \cite{Folma02} are
ideally suited to provide this field, as large field amplitudes can be created by oscillating currents of relatively low amplitude in the chip wires. Additionally, the precise control over the resulting fields and the large field gradients of wire traps allow to fully exploit the possibilities of RF potentials.

In our experiment we use a three-wire atom chip design shown in
Fig.~\ref{Fig:Setup}. A central Z-shaped wire together with an
external bias field forms a highly anisotropic Ioffe trap,
while the two wires on the sides carry oscillating currents with
frequency $\omega_{RF}$. If the static trap is placed at a
position from the surface equal to half the distance between the
RF wires, the field configuration of two perpendicular RF-fields
as discussed above is realized in the vicinity of the trap center.
By controlling the phase shift and the amplitudes $B_{A}, B_{B}$
of the oscillating fields, we gain full control over the magnitude
and polarization of the resulting RF dressing field.

Our experiments are performed with Bose-Einstein condensates (BECs)
of $10^3$ to $10^5$ $^{87}$Rb atoms. We have $\mu \simeq \hbar \omega_{\perp}$, where $\mu$ is the chemical potential of the BEC, so that we are working in the crossover regime between elongated 3D BEC and 1D-quasi condensates \cite{Petro00, Menot02}.

\section{Linear RF polarization}

\begin{figure}[tb]\center
    \includegraphics[angle=0,width=\columnwidth]{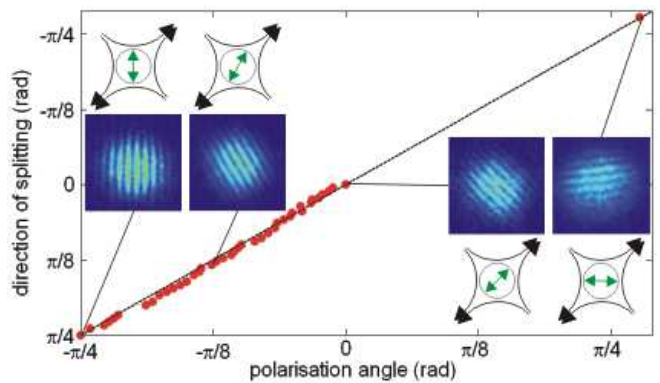}
    \caption{Orientation of the double well potential as a function of the plane of polarization in the case of a linearly polarized RF field (angles are measured with respect to the x-axis). The double well direction is inferred from interference patterns. The observed fringes are always perpendicular to the double well axis. Coherent splitting is possible in every direction.
    }\label{Fig:LinPol}
\end{figure}

We first discuss the case of a linearly polarized RF dressing
field ($\delta = 0,\pi $). The orientation of the polarization is determined by the amplitude ratio $\frac{B_B}{B_A}$. From Eq. (\ref{Eq:ring_dw2}) it follows that for any orientation, always a double well forms, with the minima located along the direction where the quadrupole component of the static magnetic field is parallel to the RF-field. The dependence of the double well orientation on the field polarization is shown in Fig. \ref{Fig:LinPol}. We study the double well rotation by observing interference patterns in time-of-flight (ToF) between the matter-wave packets released from the two wells \cite{Schum05}.

When rotating the double well, gravity adds an asymmetry of $\Delta
U = 2.1389$ [kHz/$\mu$m]$\;  \sin \phi$ to the potential. This
imbalance quickly becomes strong enough to cause the BEC being
trapped solely in the lower well. Symmetric splitting can be
achieved by bringing the double well closer to the chip surface.
The inhomogeneity of the wire fields also introduces a well controllable asymmetry when shifting the trap location by a few $\mu$m, which allows to precisely counteract the effect of gravity.
We have observed coherent splitting (the interference shows a stable phase) for all orientations of the double well.

By applying a periodic modulation to the RF amplitudes, such a
setup can be used to rotate a BEC on a circular path. The rotation
is always centered perfectly on the center of the static trapping potential.

\section{Arbitrary polarization: State dependent manipulation}

\begin{figure}[tb]\center
    \includegraphics[angle=-90,width=8cm]{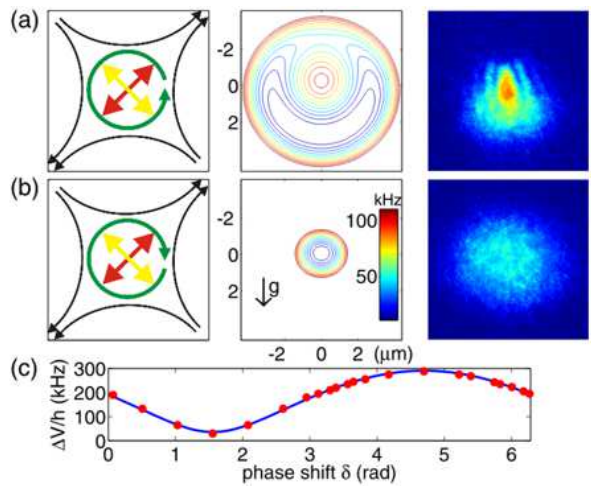}
    \caption{(a,b) RF-potentials created by circular polarization
    for Rb in the $F=2, m_F=2$ state. Depending on the handedness of the polarization \emph{(left column)} two different potentials result \emph{(center column)}. The potentials shown
    are numerical calculations for the exact wire geometry used in the experiments, including gravity. The \emph{(right column)} shows the ToF images of BECs released from either
    configuration.The peculiar interference pattern caused by the
    deformed ring is clearly visible.
    \emph{(c)} Measurement of the energy difference between dressed
    states at the trap bottom for elliptical polarization ($B_B = 1.2 \times B_A$) as a function of the phase shift $\delta$. The line is a numerical calculation for the actual experimental parameters.
    }\label{fig:CircPol}
\end{figure}

A second case of interest is that of circular (or in general
elliptical) polarization. For $B_A =B_B $ a phase shift of $\delta=\frac{\pi}{2}$ ($\frac{3\pi}{2}$) leads
to right-(left) handed circular polarization. From
Eq. (\ref{Eq:ring_dw2}) one sees that the resulting potentials
differ for either case. Depending on the sign of the g-factor
one finds a harmonic trap ($\textrm{sgn}(g_F)=+1$ (-1),
$\delta=\frac{\pi}{2}$ ($\frac{3\pi}{2} $)) or a ring potential
($\textrm{sgn}(g_F)=+1$ (-1), $\delta=\frac{3\pi}{2}$
($\frac{\pi}{2}$)).

It is interesting to note here that $\textrm{sgn}(g_F)$ defines an
\emph{effective} polarization felt by the atom. In particular, an
atom in a hyperfine state with $\textrm{sgn}(g_F)=-1$ will see the
potential formed by the opposite handedness compared to an atom with $\textrm{sgn}(g_F)=+1$. Consequently, potentials created by
circularly, or more generally elliptically, polarized RF can be
used for a state-dependent manipulation of trapped atoms.

Fig. \ref{fig:CircPol} illustrates an example of manipulating Rb
in the $F=2, m_F=2$ state with circular polarized RF. A phase shift of $\delta=\frac{3\pi}{2}$ (Fig.~\ref{fig:CircPol}a) gives a ring
shaped potential, while in the case of $\delta=\frac{\pi}{2}$
(Fig.~\ref{fig:CircPol}b) the harmonic shape of the static potential is unmodified by the RF-dressing. In the calculation of the resulting potentials as shown in Fig.~\ref{fig:CircPol} gravity has been included, and hence the ring is deformed. In our current setup, this deformation can not be fully compensated by changing the static trap and the RF parameters. Nevertheless the time-of-flight images of atoms released from the potentials illustrate the effect caused by the different handedness of the polarization. For $\delta=\frac{\pi}{2}$ the image is identical to that of
atoms released from the static trap, while for $\delta= \frac{3\pi}{2}$ we observe an interference pattern which is consistent with numerical simulations of the expansion from the ring
potential deformed by gravity. For an ideal ring shaped BEC at T=0
one would expect an interference pattern reminiscent of the
poisson spot in optics, with the size of the peak depending on the
nonlinear interaction caused by the atom-atom repulsion.

To achieve a perfect ring one would either have to change the
orientation of the atom chip resulting in a ring lying in a plane orthogonal to gravity or apply an additional field gradient canceling gravity. This can be achieved for example by an electric field gradient \cite{Krueg03}.

The effect of the polarization of the RF field on the dressed
state potentials can be directly measured by RF spectroscopy of
the trapped atoms (Fig.~\ref{fig:CircPol}c). Thereby we irradiate
the dressed BEC with a weak additional RF field. If this RF field
is resonant with the energy difference between dressed state
levels, atoms are transferred to un-trapped states, and one
observes trap loss. We measure the shift of the dressed state potential minimum as a function of the phase shift $\delta$ for the case $B_B
= 1.2 \times B_A$. The measured potential is in perfect agreement
with a numerical calculation based on the actual atom chip
dimensions. The dressed state shift varies from $\sim 50$ kHz at
$\delta = \frac{\pi}{2}$ to $\sim 320$ kHz at $\delta =
\frac{3\pi}{2}$.

\section{Advantages in designing Potentials}

To illustrate the power of RF induced adiabatic potentials for
atom manipulation we compare the splitting of a single trap into a
double well by RF to the widely discussed splitting based on a 2-wire configuration employing static fields only \cite{Casse00,Hinds01,Haens01a,Folma02,Shin05}.

\begin{figure}[tb]\center
    \includegraphics[angle=-90,width=\columnwidth]{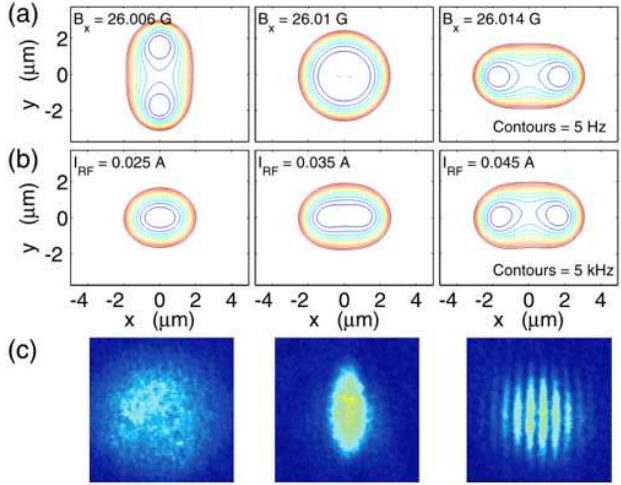}
    \caption{Comparing double well potentials created with static fields and RF dressing keeping the same structure size and distance to the chip.
    (a) 2-wire beam splitter: The splitting commences from two incoming guides evolving into two outgoing guides, through a region with a hexapol confinement. Equipotential lines are drawn at steps of 5 Hz.
    (b) Double well created by RF dressing a single Ioffe trap. The splitting transforms a single trap to a double well. The transverse confinement stays nearly constant and harmonic. At the splitting point one finds a $x^4$ confinement in the direction of the splitting. Equipotential lines are drawn at steps of 5 kHz.
    (c) ToF pictures of pure BECs released during
    the splitting process. \emph{left}: Single BEC from the original static trap.
    \emph{center}: BEC released at the splitting point.
    \emph{right}: Interference pattern from the split condensate.
    The transverse width of the ToF expansion stays constant,
    illustrating the nearly constant transverse confinement.
    At the splitting point the narrow width of the
    ToF image illustrates the reduced confinement
    in the direction of the splitting.
    }\label{Fig:compare_splitting}
\end{figure}

First one observes a drastic difference in the topology of the
splitting (Fig.~\ref{Fig:compare_splitting}): In the RF case a
single trap smoothly splits into a double well, while in the two
wire setup two incoming ports merge and split into two outgoing
ports. This potential configuration of the two wire beam splitter
\cite{Stick03} and the weak confinement in the splitting region as
discussed below are fundamental consequences of Maxwell's
equations which constrain the freedom of designing static magnetic
potentials \cite{Davis02}.

Secondly the overall confinement of the trap during the splitting
behaves very differently in either configuration. In the case of
static fields the splitting and merging process relies on higher
order multi-poles and results in a significantly weaker
confinement in \emph{both} transverse directions than in the
quadrupole based Ioffe traps at the start and the end of the
splitting sequence. In contrast, for the RF potentials the
confinement orthogonal to the splitting plane stays nearly
constant as can be seen from the ToF images in
Fig.~\ref{Fig:compare_splitting}c.

The difference between the two splitting methods can best be
quantified by approximating the symmetric double well by a
generic polynomial potential of the form
\begin{equation}
    V_\textrm{DW}=  b x^2 + d x^4
    \label{Eq:genDW}
\end{equation}
where the coefficients $b$ and $d$ determine the potential shape.
If $b>0$ there is only one minimum, for $b<0$ a double well is
formed. The overall confinement of the double well potential along the splitting direction depends only on the coefficient of the $x^4$ term ($d$). By Taylor expanding the potentials created by the RF-beamsplitter and the two-wire setup one finds
\begin{eqnarray}
  \frac{d_\textrm{RF}}{d_\textrm{2w}} \approx \left(\frac{B_\textrm{bias}}{B_\textrm{Ioffe}}\right)^2
  \label{eq:x4comarison}
\end{eqnarray}
which is on the order of $10^3$ in typical atom chip experiments.
This means that the achievable potential modulation is 1000 times
stronger in RF designed potentials than in potentials designed
with static fields alone.

Fig. \ref{Fig:compare_splitting} compares the RF double well
discussed in this paper with a 2-wire setup utilizing the same wire
dimensions and atom-chip distance. For a typical separation of $3 \mu$m and $B_\textrm{bias}=30$ G, $B_\textrm{Ioffe}=1$ G, we obtain
$\frac{d_{RF}}{d_{2w}} = 1260$, in good agreement with the
approximation (Eq.~\ref{eq:x4comarison}).

In addition, one sees that the parameter characterizing the two
wire splitting (the external bias field $B_{bias}$) has to be
stable to a precision of $10^{-5}$ for controlled splitting, whereas in the case of RF induced potentials the control parameter (the RF field amplitude) changes by a factor two, greatly reducing the effect of technical fluctuations. Altogether this demonstrates that it is significantly simpler to design and control a miniaturized
potential by using RF dressed states.

\section{Collisional stability}

A notable feature of the RF induced adiabatic potentials, as
applied in our experiments, is that all the involved frequencies,
namely the Larmor frequency  $\omega_\textrm{L}= \mathbf \mu \cdot
\mathbf{B} / \hbar$, the radio frequency $\omega_\textrm{RF}$, and
the resulting Rabi frequency $\Omega_\textrm{R}$, are of the same
order of magnitude ($\omega_\textrm{L} \sim \omega_\textrm{RF}
\sim \Omega_\textrm{R}$). In this regime the dressed eigenstates
contain contributions of all bare states defined by the
static field. Mixtures of different magnetic spin states usually
lead to high loss rates for trapped atoms \cite{Moerd96b,Julie97}.
Nevertheless we observe no additional loss for atoms trapped in RF
potentials compared to the static trap for densities up to
10$^{15}$ atoms/cm$^3$.

We attribute this collisional stability of atoms in the RF dressed
states to an argument given in \cite{Moerd96c,Suomi98} for weakly
dressed atoms during RF-evaporation in magnetic traps: The
RF-fields introduce a variation of the potential on length scales
smaller than the original static trap, but still large compared to
the scattering length of the atoms. Two colliding atom reside in the same adiabatic state, consequently they can be regarded to be in the same spin state when they approach each other closely. The RF dressing only rotates the (local) quantization axis, without affecting the collisional properties. Our observations suggest that this argument is still valid in our regime of $\omega_\textrm{L} \sim \omega_\textrm{RF} \sim \Omega_\textrm{R}$.

In contrast to the observations by \cite{Colom04} and
\cite{White06} we observe no additional heating processes in the
RF dressed state potentials.

\section{Evaporative Cooling}

With the long lifetimes, no additional heating, and high collision
rates, evaporative cooling to quantum degeneracy in the RF-induced
potentials is possible. This is accomplished by applying a sweep
of an additional weak RF field tuned to energy differences in the
dressed state level scheme. The observed RF induced evaporative
cooling is as stable and efficient as in the original magnetic
trap.

\begin{figure}[tb]\center
 \includegraphics[angle=-90,width=\columnwidth]{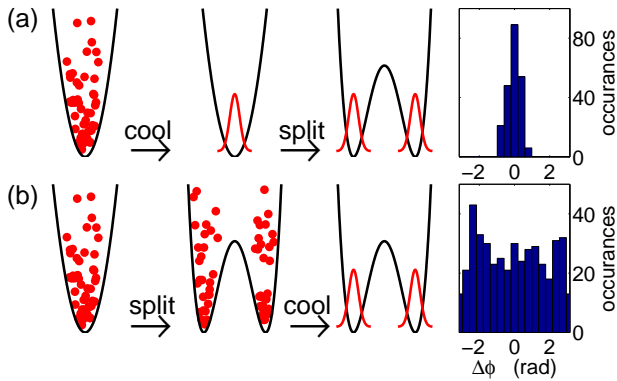}
  \caption{Comparison of independent and coherently split BECs.
  (a) For the coherent splitting a BEC is produced in the single well,
  which is then deformed to a double well. We observe a narrow phase
  distribution for many repetitions of an interference experiment
  between these two matter waves, showing that there is a
  deterministic phase evolution during the splitting.
  (b) To produce two independent BECs, the double well is formed while
  the atomic sample is thermal. Condensation is then achieved by
  evaporative cooling in the dressed state potential. The observed
  relative phase between the two BECs is completely random, as expected
  for two independent matter waves.
    \label{Fig:cool_and_split}}
\end{figure}

The ability to cool atomic samples in the RF-double well potential
allows us to prepare two fully independent condensates from a
split thermal cloud. We compare the relative phase between such BECs with that of coherently split BECs in the same potential. Fig. \ref{Fig:cool_and_split}
illustrates the two different experimental sequences. In both
cases we start with thermal atoms in the static single well
potential. In the case of coherent splitting, the sample is first
cooled to degeneracy and then the potential is deformed into a
double well by the RF dressing field
(Fig.~\ref{Fig:cool_and_split}a). Alternatively, we apply the RF
field first, splitting the thermal cloud, and then cool the two
separate samples (Fig.~\ref{Fig:cool_and_split}b).
The height of the potential barrier between the two wells is set to $U \approx 4$ $\mu$K, ensuring that the two samples
are already separated at temperatures well above the condensation temperature $T_C \approx 0.4$ $\mu$K.

We obtain the relative phase $\phi_{rel}$ between the two BECs from
the interference patterns observed in ToF after release from the
double well. For the coherently split BECs we observe a narrow
gaussian distribution of the relative phase ($\sigma = 0.2 rad$)
centered around $\phi_{rel}=0$. The independently created BECs show
the expected completely random phase distribution \cite{Casti97,Andre97a}.

In future experiments precise control over the coupling of the two 1d condensates with a tunable tunnel barrier between the two wells will allow detailed studies of the coherence dynamics like phase locking, phase diffusion and phase fluctuations and their dependence on the initial state (coherently split or independently created) and the reduced dimensionality of the BECs.

\section{Conclusion}

We have presented RF-induced dressed state potentials and
illustrated their remarkable features in experiments utilizing a
simple and highly integrated setup on an atom chip.

The RF-potentials greatly enhance the flexibility and robustness
of atom manipulation. Detailed control over the polarization
allows to create internal state dependent potentials with RF
coupling. Applying this to the $F=1, m_F=-1$ and the $F=2, m_F=1$
clock states in $^{87}$Rb, which have the same magnetic moment $\mu = \mu_B g_F m_F$ but opposite sign g-factors ($g_1 = -g_2$) will
enable high precision state dependent manipulation for QIPC
applications \cite{Calar00}. Implementation of
multi frequency RF fields can additionally
enhance the freedom to design potentials \cite{Court06}.

Furthermore we found no adverse effects of RF dressing, neither in
the collisional stability and trap loss, nor in heating. This
allows the application of RF evaporative cooling and fast and
efficient creation of BECs in the dressed state potentials.

The robustness of the RF dressing related to
collisional losses opens up new avenues to manipulate the interaction
between atoms (molecules) by tuning to molecular states and
dressing inside the collision.  With the large coupling strength
and gradients obtainable on atom chips, the formation of RF and MW
induced Feshbach resonances \cite{Moerd96c} become feasible.

From our experiments we are convinced that RF dressed state
potentials with their robustness and flexibility will very quickly
become a standard tool for cold atom manipulation comparable to
magnetic trapping or optical dipole potentials.

\paragraph{\textbf{Acknowledgement}}
We would like to thank Thorsten Schumm and Peter Kr\"{u}ger for
stimulating discussions and critical reading of the manuscript. The atom chip used in this experiment was
fabricated at the Weizman Institut of Science by S. Groth. We
acknowledge financial support from the European Union, through the
contracts IST-2001-38863 (ACQP), MRTN-CT-2003-505032 (Atom Chips),
Integrated Project FET/QIPC "SCALA" , and the Deutsche
Forschungsgemeinschaft, contract number SCHM 1599/1-1.

\section{Methods}

\paragraph{\textbf{Calculation of the dressed state potentials}}
The magnetic fields given in Eq. (\ref{Eq:Bfields}) represent a
good approximation of the fields generated by the actual
experimental setup which is shown in Fig. \ref{Fig:Setup}. For all
quantitative calculations presented here we have numerically
calculated the magnetic fields. In the course of the latter we
have taken into account the actual layout and dimensions of the
wires which essentially result in imperfections of the magnetic
quadrupole and inhomogeneities of the RF fields. The Rabi
frequency $\Omega_\text{RF}(\mathbf{r})$ and the dressed state
potentials have then been calculated following Refs.
\cite{Lesan06,Lesan06b}.

\paragraph{\textbf{Experimental Procedure}}
We routinely prepare samples of $\sim 3\times 10^{5}$ $^{87}$Rb
atoms in the F=m$_{F}$=2 state at a temperature of $\sim 2\mu$K in
the chip trap formed by the central Z-wire following our standard
procedure \cite{Wilde04}. Pure condensates containing between
$10^{3}$ and $10^{5}$ atoms are produced by a final stage of
evaporative cooling.  The number of atoms in the condensate can be
adjusted by the end frequency of the final RF cooling ramp.

The distance from the chip surface where the field configuration
given in Eq. \ref{Eq:Bfields} is realized is $115$ $\mu$m. At this
height residual potential roughness has not been observed with
even the most sensitive measurements \cite{Wilde05b, Krueg05} and
is extrapolated to be many orders of magnitude smaller than the
chemical potential of the BEC.

The transverse confinement in the static trap can be adjusted by
the wire current and the external bias field. The parameters are
chosen such that $\omega_{\perp} = 3$ kHz. The longitudinal
confinement is provided by the outer leads of the Z-shaped wires,
in addition it can be modified by the U-shaped wires on the side
of the three-wire setup (Fig.~\ref{Fig:Setup}b). For the
experiments described here, we used $\omega_{\parallel} \sim 10$
Hz. The chemical potential $\mu$ of the BEC in this trap fulfills
$\mu \simeq \hbar \omega_{\perp}$, so that we are working in the
1D crossover region.

\paragraph{\textbf{Imaging the atoms}}
The atomic clouds are studied either \emph{in situ} or in time-of-flight by resonant absorption imaging. The diffraction limited resolution of the imaging system is adjusted to have a focal depth compatible with the longitudinal size of the atomic cloud. In the experiments here it was set to $\sim 5$ $\mu$m to allow imaging of 100 $\mu$m long clouds.

\paragraph{\textbf{RF spectroscopy of dressed state potentials}}
The energy difference between different dressed state potentials
can be measured by inducing transitions with a weak additional RF
field, transferring atoms to untrapped states. This induces trap
loss, which is the signature for achieving resonance. The
resolution of this method is limited by the chemical potential of
the BEC to $\sim 2$ kHz.

\paragraph{\textbf{Evaporative cooling in dressed state potentials}}
One can use the loss caused by transitions between dressed state
energy levels for evaporative cooling. Unlike for the case of a
static magnetic trap not only one but multiple resonance
frequencies between the dressed energy levels at
which atoms can be transferred to un-trapped states can be identified. We find that the efficiency of the evaporation process differs for these resonances.

\end{document}